\documentstyle[12pt,leqno]{article}
\def\hs#1{\hskip #1pt}
\def\vs#1{\vskip #1pt}
\def\smalldisplayskip{\noalign{\vskip 5pt plus 1pt minus 2pt}}
\def\ss{\scriptstyle}
\def\sss{\scriptscriptstyle}
\def\llongrightarrow{\relbar\mkern-10mu{\relbar\mkern-10mu
{\relbar\mkern-10mu{\longrightarrow}}}}
\def\eq#1{\hbox{eq.\hs{2}(#1)}} 
\def\eqs#1{\hbox{eqs.\hs{2}(#1)}}
\def\Eq#1{\strut\hbox{Eq.\hs{2}(#1)}} 
\def\fig#1{\hbox{fig.\hs{2}#1}} 
\def\ie{\strut\hbox{i.\hs{2}e.} } 
\def\iec{\strut\hbox{i.\hs{2}e.}} 
\def\eg{\hbox{e.\hs{2}g.} } 
 
\def\R{{\rm I\! R}}
\begin{document}
\title{Quantum measurement\\
in a family of hidden--variable theories\thanks{Supported
in part by the Istituto Nazionale di Fisica Nucleare.}}

\author{Giulio Peruzzi and Alberto Rimini\\
{\em  Dipartimento di Fisica Nucleare e Teorica,\/}\\ 
{\em Universit\`a di Pavia}}
\date{June 28, 1996}
\maketitle

\begin{abstract}
The measurement process for hidden--configuration formulations of quantum 
mechanics is analysed. It is shown how a satisfactory description of quantum 
measurement can be given in this framework. The unified treatment of 
hidden--configuration theories, including Bohmian mechanics and Nelson's 
stochastic mechanics, helps in understanding the true reasons why the problem 
of quantum measurement can succesfully be solved within such theories. 
\end{abstract}

\section{Introduction}

According to the so--called Copenhagen interpretation supporting the standard 
formulation of quantum mechanics, it is admitted that we can identify two 
different situations: the undisturbed evolution of a (quantum) micro--system 
(${\cal S}$) and the measurements, \ie the interactions of $\cal S$ with 
suitable (classically described) macroscopic devices ($\cal A$). 
In the standard formulation, the ordinary time evolution of the state of 
$\cal S$ is governed by the Schr\"{o}dinger equation. This deterministic 
evolution is suspended when a measurement is performed. The acts of 
measurement are governed via special rules ({\it probability rule\/} and 
{\it reduction postulate\/}) having a stochastic character. Measurements have 
usually many possible outcomes and in general the knowledge of the state of 
the system {\it before the measurement\/} does not determine the actual 
outcome but only a probability distribution for the various different outcomes 
({\it probability rule\/}). The state of the system {\it after the 
measurement\/} is determined by the actual outcome ({\it reduction 
postulate\/}). The a priori acceptance of the dualism of the physical 
situations is logically a necessary condition for the acceptance of the 
dualism of the evolution principles. 

One of the big difficulties connected with such forms of dualism in quantum 
mechanics is that the distinction between the quantum system and the 
classically described measuring apparatus, even though easy in practice, is 
not sharply defined in principle and this introduces a basic vagueness into 
the fundamental physical theory. The theory of quantum measurement tries to 
get rid of the dualism assuming the Schr\"{o}dinger equation as the sole 
principle of evolution for the composite system ${\cal S} + {\cal A}$.
We shall briefly review, in Section 2, the hard difficulties met by this 
programme, giving also a r\'esum\'e of the families of theories which overcome 
these difficulties without loosing anything as regards the description of the 
quantum--mechanical phenomenology. The hidden--configuration theories, which 
include de Broglie--Bohm's theory \cite{Bohm} and Nelson's stochastic 
mechanics \cite{Nelson,Nelson1}, are one of such families. As described in 
Section 3 some results of Davidson \cite{Davidson} and of Bohm and Hiley 
\cite{Bohm-Hiley1} allow to present in a unified way the whole family. The 
measurement process in this framework will be analysed in Section 4. In the 
last Section we shall present some concluding remarks.

The theory of quantum measurement in the de Broglie--Bohm theory is due to 
Bell \cite{Bell,Bell1} and Bohm and Hiley \cite{Bohm-Hiley}. Recently, it has 
been revisited by D\"{u}rr, Goldstein and Zangh\`{\i} \cite{DGZ} in the 
framework of Bohmian mechanics, and by Rimini \cite{Rimini} through a simple 
but significant example. 

On the other side, the situation of the theory of quantum measurement in the 
stochastic formulation of quantum mechanics is more involved. Whereas 
Blanchard, Golin and Serva \cite{BGS} consider the problem of quantum 
measurements without including the apparatus $\cal A$, the stochastic 
mechanical treatment of the whole ${\cal S} + {\cal A}$ is given by Goldstein 
\cite{Goldstein} and Blanchard, Cini and Serva \cite{BCS}. It is stated, in 
both the latter papers, that the solution of the problem of quantum 
measurement in stochastic mechanics is based on the decoherence of the wave 
function and on the role of hidden configuration variables. While we 
completely agree with the treatment of Goldstein, we find some interpretative 
uncertainties in the point of view of Blanchard, Cini and Serva. 

A totally different approach, based on the theory of stochastic variational 
principles and their discrete generalizations, is described by Guerra 
\cite{Guerra}. Starting from a single variational principle, one can simulate, 
together with critical processes reducing to the usual ones of stochastic 
mechanics (in the continuous limit), also processes simulating the formation 
of a mixture, in analogy with the quantum measurement collapse. The physical 
implications for the theory of quantum measurement, and, before that, a 
physical justification for the assumed form of the stochastic Lagrangian, are 
still open problems. 

In the present paper, we adopt a unified formulation of hidden--configuration 
theories and show that a satisfactory treatment of quantum measurement can be 
given in the framework of such a formulation. We think that the unified 
treatment helps in understanding the true reasons why the problem of quantum 
measurement can succesfully be solved within such theories.

\section{The problem of quantum measurement}

Let us consider the measurement process including the measuring apparatus 
${\cal A}$ together with the measured system ${\cal S}$ in the description. 
Different outcomes of the measurement correspond to different final pointer 
positions and consequently to different, macroscopically distinguishable 
states of ${\cal A}$ and ${\cal S} + {\cal A}$. 

In the standard formulation the state is identified with the state vector and 
the correspondence between the outcome and the state is enforced by the 
reduction postulate. Let ${\Psi}^{\rm i}$ and ${\Psi}^{\rm f}$ be the state 
vectors of ${\cal S} + {\cal A}$ before and after the measurement, 
respectively. For definiteness and simplicity let us consider ideal 
measurements (this is not essential for the arguments which follow). The 
evolution from ${\Psi}^{\rm i}$ to ${\Psi}^{\rm f}$ will be of the form
\begin{equation}
{\Psi}^{\rm i}=\sum_{m}c_{m}{\varphi}_{m}{\alpha}^{\rm i} 
~~~\llongrightarrow~~~{\Psi}^{\rm f} = \cases{
{\varphi}_{1}{\alpha}^{\rm f}_{1}, \hs{12} 
\hbox{with probability $|c_{1}|^{2}$}\cr
\smalldisplayskip
{\varphi}_{2}{\alpha}^{\rm f}_{2}, \hs{12} 
\hbox{with probability $|c_{2}|^{2}$}\cr
\smalldisplayskip
\cdots \cdots \hs{12} \cdots \cdots \cdots \cdots \cdots \cdots \cdots}
\end{equation}
with obvious meaning of symbols. The output of the measurement is not uniquely 
determined, that is the final situation is a mixture represented by the 
statistical ensemble made up with the various outputs with their proper 
weights. This is the consequence of the probability rule. When the measurement 
is completed, we will get one and only one definite outcome which corresponds 
to a definite state of ${\cal S} + {\cal A}$. If we immediately repeat the 
same measurement on the same system we shall obtain with certainty (\ie with 
probability equal to one) the same outcome. 

The program of the theory of quantum measurement, however, is to assume the 
Schr\"{o}dinger equation as the sole principle of evolution for the composite 
system ${\cal S} + {\cal A}$. Because of the linearity of the Schr\"{o}dinger 
equation, the evolution from ${\Psi}^{\rm i}$ to ${\Psi}^{\rm f}$ will be of
the form 
\begin{equation}
{\Psi}^{\rm i}=\sum_{m}c_{m}{\varphi}_{m}{\alpha}^{\rm i}
~~~{\buildrel {\rm S.E.} \over \llongrightarrow}~~~
{\Psi}^{\rm f}=\sum_{m}c_{m}{\varphi}_{m}{\alpha}^{\rm f}_{m}.
\end{equation}
The final state ${\Psi}^{\rm f}$ in \eq2 is a superposition of macroscopically 
distinguishable states, a pure state, not a mixture. In such a state all 
outcomes coexist, no matter of the fact that $\cal S + \cal A$ is macroscopic. 
This {\it inconceivable state\/} is the rub leading to the familiar paradoxes 
associated with quantum measurement, such as Schr\"{o}dinger's cat 
\cite{Schroedinger} or Wigner's friend \cite{Wigner}.

A first family of attempts to overcome these difficulties is provided by 
theories based on what can be called {\it ensemble interpretation\/}. This 
consists essentially in the following three statements:

\vs6
\noindent (A) the state vector ${\Psi}$ describes an ensemble of identically 
prepared physical systems, (it does not describe an individual system);

\vs6
\noindent (B) the state vector ${\Psi}$ always evolves according to the 
Schr\"{o}dinger equation;

\vs6
\noindent (C) the link between the state vector and the results of experiments 
is given by the standard probability rule.

\vs6

\noindent
These theories can be seen as the most immediate extensions of standard 
quantum mechanics, in that they develop the program of the theory of quantum 
measurement referring only to the standard formalism plus some extra 
assumptions and/or some crucial remarks. 

The goal of the theories of measurement based on the ensemble interpretation 
is the proof that the final situations in (1) and (2) are equivalent. Before 
proceeding, we must make statement (C) more precise. Let us assume that the 
probabilities to which this statement refers are the probabilities of the 
outcomes of any (realized or realisable) measurement on the considered 
physical system, at any chosen instant of time. 
equivalence of the final situations in (1) and (2) means proving that the 
outcomes of any further measurement realisable on the ensemble of the systems 
${\cal S} + {\cal A}$ are the same in the two cases. This, clearly, involves 
the need of suitable limitative assumptions on the measurements which can 
actually be performed. Different limitative assumptions correspond to 
different theories in the family of those based on the ensemble 
interpretation. Starting with different limitative assumptions, all these 
theories, in practice, converge in the proof that the interference effects 
between different terms in (2) cannot be experimentally revealed. Such a 
result can be called {\it effective incoherence\/}. We note that, in this type 
of approach, even though the measurement apparatus is included in the 
description, one does not avoid the reference to further measurements to be 
performed at successive times. Therefore the vague concept of ``measurement'' 
is not removed at the level of principles, and this fact is detrimental to the 
claim of descriptive precision.

A particular subfamily of theories of measurement based on the ensemble 
interpretation makes use of the pratically inescapable interaction of 
macroscopic systems with their environment, $\cal E$ (cf. the work of Joos and 
Zeh \cite{Joos-Zeh} and the works of Gottfried \cite{Gottfried}). In this case 
the limitative assumption consists in the recognition of the practical 
inaccessibility by means of a measurement of all the degrees of freedom of 
$\cal E$. This is the weakest limitation, in the sense that no assumption on 
the measurability of $\cal S + \cal A$ is necessary. For this reason, we make 
reference in the subsequent discussion to this approach, having in mind that, 
the same problems of descriptive precision and, as we shall see, of 
consistency arise anyhow. The evolution in \eq2 is then replaced by 
\begin{equation}
{\Psi}^{\rm i}=\sum_{m}c_{m}{\varphi}_{m}{\alpha}^{\rm i}{\eta}^{\rm i}
~~~{\buildrel{\rm S.E.} \over \llongrightarrow}~~~{\Psi}^{\rm f}=
\sum_{m}c_{m}{\varphi}_{m}{\alpha}^{\rm f}_{m}{\eta}^{\rm f}_{m},
\end{equation}
where $\eta$ denotes the states of $\cal E$ (or of its part inaccessible to 
measurement). If the states ${\eta}^{\rm f}_{m}$ are and remain practically 
orthogonal, the effective incoherence of the different terms in 
${\Psi}^{\rm f}$ immediately follows. In other words, the ensemble 
corresponding to the mixture and the one corresponding to the pure state 
cannot be distinguished by measurements which do not involve also ${\cal E}$.

However, besides the problem of precision mentioned above, a problem 
of consistency arises. Let the measurement in \eq3 be performed at the time 
$t_{1}$ and the same measurement be repeated at the time $t_{2} > t_{1}$ by 
means of a similar apparatus $\cal A'$. The evolution will now be 
\begin{eqnarray}
{\Psi}^{\rm i}=\sum_{m}
c_{m}{\varphi}_{m}{\alpha}^{\rm i}{{\alpha}{\ss '}}^{\rm i}{\eta}^{\rm i} 
&  {\buildrel{\rm S.E.}\over\llongrightarrow} &
{\Psi}_{t_1}=\sum_{m}c_{m}{\varphi}_{m}{\alpha}^{\rm f}_m
{{\alpha}{\ss '}}^{\rm i}{\eta}^{\rm f}_m \\
& {\buildrel{\rm S.E.}\over\llongrightarrow} &
{\Psi}_{t_2}=\sum_{m}c_{m}{\varphi}_{m}{\alpha}^{\rm f}_m
{{\alpha}{\ss '}}^{\rm f}_m{\eta{\ss '}}^{\rm f}_m ,
\nonumber
\end{eqnarray}
with obvious meaning of symbols. The quantum probability law can then be 
applied in two different ways. First, to the state vector after the first 
measurement (${\Psi}_{{t}_{1}}$) to calculate the probability distribution of 
the second measurement. In this case, if one turns her/his the attention to 
the (selected) subensemble for which, say, the result $l$ has been obtained, 
one gets a {\it wrong answer\/}. Second, to the state vector after the second 
measurement (${\Psi}_{{t}_{2}}$) to calculate the probability of a third 
measurement consisting in the joint reading of the pointers of apparatuses 
$\cal A$ and ${\cal A}'$. In this case one gets the {\it correct answer\/} 
that the first and the second measurements gave the same result. Such an 
inconsistency is produced because in the description of the system after a 
measurement there is no trace of the outcome obtained for a specific element 
of the ensemble, or, at least, for the elements of a (selected) sub-ensemble 
of the original ensemble: every element is always described by the whole 
${\Psi}_{t}$. A similar inconsistency is obtained, even without performing 
selections, in the case of a finite ensemble \cite{Nicrosini}.

We can summarize the above discussion in the following way.
Let us consider the statement 

\bigskip

\centerline{\vbox{\hsize 300pt \noindent 
(CSR) there is a correspondence between the outcome of
a measurement and the description (state) of the system 
${\cal S} + {\cal A}$ after the measurement. 
}}

\bigskip

\noindent
Let us call this proposition the {\it Common Sense Requirement\/} (CSR). The 
standard formulation, including the reduction principle, satisfies CSR. On the 
other hand, if we pretend that Schr\"{o}dinger's equation be the sole 
principle of evolution, we get the inconceivable state, \ie a description of 
${\cal S} + {\cal A}$ after the measurement in which there is {\it no 
counterpart\/} of the individual outcomes, so that CSR is not satisfied. 
Adding limitative assumptions on measurability does not make CSR satisfied, 
causing the type of inconsistency discussed above. 

Statement (C) can be made precise also in a way different from that proposed 
above. Let us assume that the probabilities to which statement (C) refers are 
the probabilities that the situations described in the various terms in the 
wave function $\Psi_t$ (say the wave functions appearing in \eq4) come true. 
At first sight, this new interpretation could appear a slight modification of 
the previous one. On the contrary, in this way, statement (A) is radically 
changed: the pair $({\Psi}, {m})$ --- the wave function and the branch label, 
\ie the specification of the term in the wave function --- describes an 
individual system. One may question whether such an additional element of 
description is introduced in a satisfactory way, but, as a matter of fact, the 
branch label is there and it is used to describe individual systems. Both 
problems of precision (no reference to the concept of measurement is there at 
the level of principles) and consistency  (in fact CSR is satisfied) are 
solved, but we are far away the conceptual framework of the ensemble 
interpretation. Actually, specifying the statement (C) as above leads to the 
so called {\it consistent histories\/} interpretation of quantum mechanics 
\cite{Griffiths,Omnes,Gell-Mann}. The details of such an interpretation are 
beyond the aim of this paper.

There are two other ways of recovering CSR maintaining a single principle of 
evolution: {\it hidden variables theories\/} and {\it reduction theories\/}. 

In hidden variables theories, like the de Broglie--Bohm theory and Nelson's 
stochastic mechanics, the evolution of $\Psi$ is always governed by the 
Schr\"{o}dinger equation, but, contrary to the standard formulation, $\Psi$ 
does not exaust the description of the system. A new variable is added to 
$\Psi$: this (hidden) variable evolves in such a way that its value after the 
measurement is in correspondence with the outcome. 

In reduction theories, like the GRW theory \cite{GRW}, the evolution of 
$\Psi$, which completely specifies the state, is changed assuming a new 
principle of evolution. This is essentially equivalent to the Schr\"{o}dinger 
equation in ordinary situations but it incorporates reduction in the 
measurement situations: the state vector collapses stochastically in 
accordance with quantum mechanical probabilities.

\section{Hidden--configuration theories}

In hidden--configuration theories, the state of an $N$--particle system at 
time $t$ is described by the pair $(\mbox{\boldmath $x$}, \Psi)$, where 
$\mbox{\boldmath $x$}(t)=({x}_{1}(t), ...,{x}_{N}(t))$ is the point in 
configuration space, ${\R}^{3N}$, and $\Psi(\mbox{\boldmath $x$},t)$ is a 
vector in the quantum Hilbert space. If the configuration of the system or of 
part of it is measured at a certain time, the system or its part are found as 
specified by the value of $\mbox{\boldmath $x$}$ at that time. The time 
evolution of the pair $(\mbox{\boldmath $x$}, \Psi)$ is given by
\begin{eqnarray}
i\hbar \frac{\partial \Psi}{\partial t} & = & H \Psi,\\
{\rm d}\mbox{\boldmath $x$} & = & \mbox{\boldmath $b$}(\mbox{\boldmath $x$},t) 
\,{\rm d}t + \sqrt{\alpha} \,{\rm d}\mbox{\boldmath $w$}. 
\end{eqnarray}
\Eq5 is the Schr\"{o}dinger equation with Hamiltonian $H$ given by
\begin{equation}
H = - \frac{{\hbar}^{2}}{2} \frac{\bigtriangleup}{m} + V
\end{equation}
(we consider for simplicity spinless particles), where 
${\bigtriangleup}/{m} = \sum_{i}{\bigtriangleup}_{i}/{m}_{i} $. For 
all regions where $\Psi$ is different from zero we can write it in the form 
\begin{equation}
\Psi(\mbox{\boldmath $x$},t) = \exp \left[R(\mbox{\boldmath $x$},{t}) + 
iS(\mbox{\boldmath $x$},{t}) \right],
\end{equation}
with $R$ and $S$ time dependent real functions on ${\R}^{3N}$. In \eq6 
$\mbox{\boldmath $w$}(\mbox{\boldmath $x$},t) = 
({w}_{1}({x}_{1},t),...,$ ${w}_{N}({x}_{N},t))$ 
is a set of $N$ independent Wiener processes in ${\R}^{3}$ such that 
\begin{equation}
\overline{{{\rm d}w}_{i}}=0, \;\;\; 
\overline{({{\rm d}w}_{i})^{2}}=\frac{\hbar}{m_{i}},
\end{equation}
and $\alpha$ is a positive real constant. Then, for any $\alpha$, 
$\mbox{\boldmath $x$}(t)$ is a Markov process in  ${\R}^{3N}$ 
characterized by the diffusion constants\footnote{In general, one can consider 
for each kind of particles ${\nu}_{i} = {\alpha}_{i}  \hbar / m_{i}$.
The results below remain valid in the general case, while the choice of a 
unique $\alpha$ simplifies the notation.}
\begin{equation}
{\nu}_{i} = \alpha \frac{\hbar}{m_{i}} 
\end{equation} 
and by the forward drift $\mbox{\boldmath $b$}= ({b}_{1},...,{b}_{N})$.
We shall not discuss the possible physical origins of these fluctuations. 
In analogy with the use in Bohmian mechanics, \eq6 can be called the {\em 
guidance equation\/}. If $\rho (\mbox{\boldmath $x$},t)$ represents the 
probability density of the stochastic process $\mbox{\boldmath $x$}$, the 
backward drift $\mbox{\boldmath $b^{\ast}$}$ is given by
\begin{equation}
\mbox{\boldmath $b^{\ast}$} =\mbox{\boldmath $b$} - \alpha \hbar
\frac{1}{\rho}\frac{\mbox{\boldmath $\nabla$}}{m}\rho,
\end{equation}
and the current density $\mbox{\boldmath $j$}
(\mbox{\boldmath $x$},t) =({j}_{1},...,{j}_{N})$ by
\begin{equation}
\mbox{\boldmath $j$}=\frac{1}{2} (\mbox{\boldmath $b$} + 
\mbox{\boldmath $b^{\ast}$}) \rho=
\mbox{\boldmath $b$}\rho - \frac{1}{2}\alpha \hbar\frac{1}{\rho}
\frac{\mbox{\boldmath $\nabla$}}{m} \rho,
\end{equation}
with  $\mbox{\boldmath $\nabla$}/{m} 
= ({\nabla}_{1}/{m}_{1},...,{\nabla}_{N}/{m}_{N})$.
$\rho$ and $\mbox{\boldmath $j$}$ satisfy the continuity equation
\begin{equation}
\frac{\partial \rho}{\partial t} + 
\mbox{\boldmath $\nabla$} \cdot \mbox{\boldmath $j$}=0.
\end{equation}

Let us assume that the probability density for the stochastic process ${\rho}$ 
is given at some initial time $t=t_{\scriptscriptstyle 0}$ by
\begin{equation}
{\rho}(\mbox{\boldmath $x$},{t})  = 
|{\Psi}(\mbox{\boldmath $x$},{t}) |^{2}
=\exp\left[ 2 R(\mbox{\boldmath $x$},{t})\right].
\end{equation}
This relation remains true at all times if the quantum continuity equation
\begin{equation}
\frac{\partial \exp\left( 2R\right)}{\partial t} + 
\mbox{\boldmath $\nabla$} \cdot \hbar\left[ \left(
\frac{\mbox{\boldmath $\nabla$}}{m} S\right) \exp\left( 2R\right) \right] = 0
\end{equation}
coincides with the continuity equation of the stochastic process (13).
This condition implies
\begin{equation}
\mbox{\boldmath $j$} = \hbar \left(
\frac{\mbox{\boldmath $\nabla$}}{m} S\right)\exp\left( 2R\right),
\end{equation}
\ie assuming,
for the forward drift $\mbox{\boldmath $b$}$ appearing in \eq6,  
\begin{equation}
\mbox{\boldmath $b$}   =  \hbar 
\frac{\mbox{\boldmath $\nabla$}}{m} \left(\alpha  R +  S \right).
\end{equation}

One may raise the objection that, as already noted just above \eq8, the 
formulation of quantum mechanics via the diffusion process described by (6) 
with $\mbox{\boldmath $b$}$ and  $\mbox{\boldmath $j$}$ given by 
(17)\footnote{Equation (17) can be also written in the form 
\[  \mbox{\boldmath $b$}   =  \frac{\hbar}{m}\left( \alpha \mbox{Re}
\frac{\mbox{\boldmath $\nabla$}\Psi}{\Psi}  + \mbox{Im}
\frac{\mbox{\boldmath $\nabla$}\Psi}{\Psi}   \right).\]} and (16) is well 
defined only for the regions of the configuration space where $\Psi$ is 
different from zero and sufficiently regular. The drift 
$\mbox{\boldmath $b$}$, for example, could have singularities on the nodal 
surfaces of $\Psi$ or in correspondence of possible singular points of the 
potential $V$, where $\Psi$ has singularities. It is proved (see 
\cite{Carlen,Nelson1} for Nelson's mechanics, and in general for quantum 
stochastic processes like those described here, and see \cite{BDGPZ} for 
Bohmian mechanics) that for a general class of potentials (including the case 
of $N$-particle Coulomb interaction with arbitrary charges and masses) the 
dynamics of hidden--configuration theories is well defined. Roughly speaking, 
the relation between $\Psi$ and the diffusion process is formulated in such a 
way that if a particle trajectory starts in the complement of the nodal set 
and of the set of singularities of $\Psi$ it does not leave this region with 
probability equal to one. As a consequence $\mbox{\boldmath $b$}$ and 
$\mbox{\boldmath $j$}$ can be considered equal to zero on the nodal set and on 
the set of singularities of $\Psi$.

Assuming that all measurements can be reduced to position 
measure\-ments,\footnote{This assumption is called ``sufficiency of position 
measurement'' by Goldstein \cite{Goldstein}. In fact, if a variable does not
commute with position, its measurement is achieved by a time-delayed position 
measurement. This has been firstly remarked by Feynman and Hibbs 
\cite{Feynman} and applied in the framework of the path integral formulation 
of quantum mechanics. The same result, in the context of the stochastic 
interpretation of quantum mechanics, can be found in Ghirardi, Omero, Rimini 
and Weber \cite{Ghirardi-Omero}.} the hidden--configuration theories and 
standard quantum mechanics are experimentally indistinguishable. In fact, 
hidden--configuration theories and standard quantum mechanics predict the same 
probabilities of finding the system in any configuration at any time, if 
${\rho} = |{\Psi}|^{2}$ is satisfied at the initial time.

Even though the presentation is completely different, the key idea of this 
section, that is the construction of a family of stochastic theories 
equivalent to standard quantum mechanics, is due, as said above, to Davidson. 
It is easily seen that for $\alpha = 0$ Bohmian mechanics and for $\alpha = 1$ 
Nelson's stochastic mechanics are obtained. 

We remark that, in the formulation adopted here, the existence of the $\Psi$ 
field which evolves according with the Schr\"{o}dinger equation is assumed ab 
initio. For such a reason, the problem of equivalence between the Madelung 
hydrodynamic equations and the Schr\"{o}dinger equation, discussed by 
Wallstrom \cite{Wall}, is bypassed. Correspondingly, any reference to quantum 
potentials or to Newton's equation is eliminated.

It is important to note, in preparation for the next section, that if $\Psi$ 
splits into two (or more) parts, $\Psi = {\Psi}_{1} + {\Psi}_{2}$, in such a 
way that the Schr\"{o}dinger evolution keeps the two terms separated in 
configuration space during a certain time, then
\begin{equation}
\rho = {\rho}_{1} + {\rho}_{2}
\end{equation}
and
\begin{equation}
\mbox{\boldmath $j$} = \mbox{\boldmath $j$}_{1} +
\mbox{\boldmath $j$}_{2},
\end{equation}
where ${\rho}_{1}$, $\mbox{\boldmath $j$}_{1}$ and ${\rho}_{2}$, 
$\mbox{\boldmath $j$}_{2}$ are nonzero, respectively, in two domains in 
configuration space ${D}_{1}$ and ${D}_{2}$ disjoint during the considered 
time. According to \eqs{14} and (16), for the purpose of the evolution of a 
system whose $\mbox{\boldmath $x$}$ lays, say, in ${D}_{1}$ the term
${\Psi}_{2}$ can consistently be dropped. A similar decomposition 
\begin{equation}
\mbox{\boldmath $b$} = \mbox{\boldmath $b$}_{1} +
\mbox{\boldmath $b$}_{2}.
\end{equation}
holds for the drift $\mbox{\boldmath $b$}$. It follows from the very general 
results quoted above the (almost sure) impossibility of the configuration of 
evolving in such a way to cross the boundaries of the disjoint supports.

\section{Theory of measurement in hid\-den--con\-fi\-gu\-ra\-tion theories}

\font\wf=cmsy10 at 25pt 
\chardef\terzultimo="7D 
\chardef\penultimo="7E
\def\quadri{{\wf \terzultimo}}
\def\cuori{{\wf \penultimo}}
\def\cpsiuno{${\ss c}_{\sss1}{\ss\psi}_{\sss1}$}
\def\cpsidue{${\ss c}_{\sss2}{\ss\psi}_{\sss2}$}
\def\psiuno{${\ss\psi}_{\sss1}$}
\def\psidue{${\ss\psi}_{\sss2}$}
\def\cpsi{\vbox{%
\vbox to 20mm{\vskip 5.9mm\hbox{\cpsiuno} \vfil}%
\vbox to 20mm{\vfil \hbox{\cpsidue} \vskip 5.9mm}}}
\def\cuoqua{\vbox{%
\vbox to 20mm{\vfil \hbox{\cuori} \vfil}%
\vbox to 20mm{\vfil \hbox{\quadri} \vfil}}}
\def\punto{\vbox to 40mm{\vskip 7.6mm \hbox{\bf .} \vfil}}
\def\puntop{\vbox to 40mm{\vskip 8.6mm \hbox{\bf .} \vfil}}
\def\rett{\hbox{
\vrule height 20mm depth 0.1mm width 0.1mm 
\vbox to 20mm{\hrule height 0mm depth 0.1mm width 2.5mm 
\vfill \hrule height 0mm depth 0.1mm width 2.5mm}
\kern-1.4mm{\vrule height 20mm depth 0.1mm width 0.1mm}}}
\def\black{\hbox{\kern 1.42mm{\vrule height 20mm depth 0mm width 2.7mm}}}
\def\numeri{\vbox to 42mm{\hbox{$\ss(1)$}%
\vfill\hbox{\lower 1mm\hbox{$\ss(2)$}}}}
\def\numerip{\vbox to 42mm{\hbox{$\ss(1{\sss'})$}%
\vfill\hbox{\lower 1mm\hbox{$\ss(2{\sss'})$}}}}
\def\freccia{\vtop{\vskip-23mm\hbox{$\buildrel \hbox{\sixrm S.E.+G.E.}\over 
\llongrightarrow$}\vfill}}
\def\inizfig{\vtop{\hsize 33mm\hbox{\hskip 2mm%
\cpsi\cuoqua\kern -15.65pt\punto\hskip11.67mm\vbox{\rett\rett}%
\hskip2pt\numeri}}}
\def\finfig{\vtop{\hsize 53mm\hbox{\hskip 17mm%
\vbox{\black\rett}\hskip2pt\numeri\hskip 3.33mm\cuoqua%
\cpsi\kern -9mm\puntop}}}
\def\inizdes{\vtop{\hsize 33mm\hbox{\vrule width 33mm height .2mm depth 0mm}%
\vskip -0.5mm%
\hbox{\hskip 7mm$\psi(x)$ \hskip 5mm $\beta^{\rm i}(\mbox{\boldmath $y$})$}%
\vskip -2mm%
\hbox{\vrule width 33mm height .2mm depth 0mm}%
\vskip -1mm%
\hbox{\hskip 9mm$\bar x$ \hskip 10.5mm $\mbox{\boldmath $y$}^{\rm i}$}%
\vskip -2.5mm%
\hbox{\vrule width 33mm height .2mm depth 0mm}}}
\def\findes{\vtop{\hsize 53mm\hbox{\vrule width 53mm height .2mm depth 0mm}%
\vskip -0.5mm%
\hbox{\hskip 2mm$c_1\psi_1(x)\beta^{\rm f}_1(\mbox{\boldmath $y$})%
\!+\!c_2\psi_2(x)\beta^{\rm f}_2(\mbox{\boldmath $y$})$}%
\vskip -2mm%
\hbox{\vrule width 53mm height .2mm depth 0mm}%
\vskip -1mm%
\hbox{\hskip 18mm$z_1,\!...$\hskip 6.5mm $\bar x'$}%
\vskip -2.5mm%
\hbox{\vrule width 53mm height .2mm depth 0mm}}}
\def\figura{\hbox{\vtop{\inizfig\vskip 1mm\inizdes}%
\hskip 14mm\freccia\hskip 2mm%
\vtop{\finfig\vskip 1mm\findes}}}

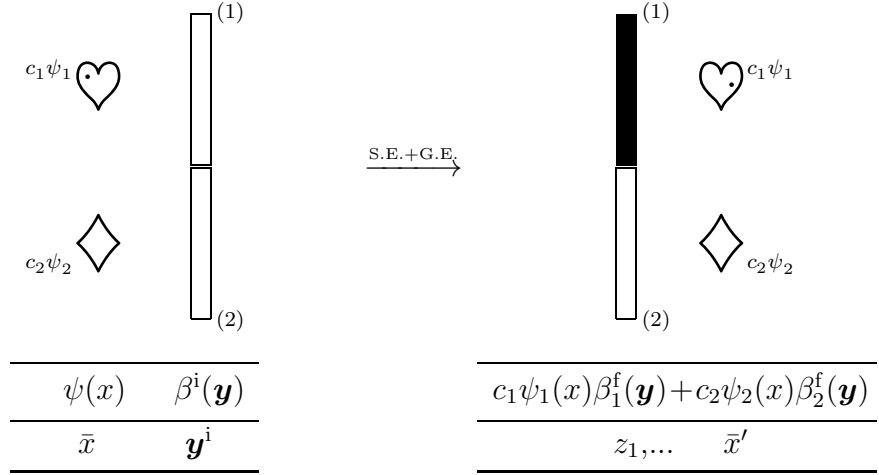
\def\dicitura{Figure 1: The evolution of the system ${\cal S}+{\cal B}$ during 
the coarse position measurement according to hidden--configuration theories. 
Between the lines at the bottom, the complete descriptions of the system 
${\cal S}+{\cal B}$ before and after the measurement are indicated. The dot 
in the term $c_{1} {\psi}_{1}$ of $\psi$ represents the actual positions 
$\overline{x}$, $\overline{x}'$ of the particle.}

We consider the quantum measurement process in the framework of 
hid\-den--con\-fi\-gu\-ra\-tion theories through a simple but significant 
example \cite{Nicrosini,Rimini}, in which a coarse measurement of the position 
of the particle $\cal S$ is performed by a pair $\cal A$ of detectors (see 
\fig1). For simplicity, we shall indicate by $\cal B$ the pair $\cal A$ of 
detectors plus the environment $\cal E$. In hid\-den--con\-fi\-gu\-ra\-tion 
theories, the state of the system ${\cal S} + {\cal B}$ is given by 
\[\left( (x,\mbox{\boldmath $y$}),\Psi (x, \mbox{\boldmath $y$}) \right) \] 
\noindent 
where $x$ is the position of the particle, $\mbox{\boldmath $y$}$ is the 
configuration of the particles in $\cal B$, and $\Psi (x,\mbox{\boldmath $y$}) 
 = \sum_i c_i\psi_i (x) \beta_i (\mbox{\boldmath $y$})$ is the wave function 
of the entire system. A subset $z$ among the coordinates $\mbox{\boldmath 
$y$}$ plays the role of pointer variable of the detectors. The value, say, 
$z= {z}_{1}$ ($= {z}_{2}$) corresponds to the first (second) detector having 
revealed the particle. Let the particle wave function before the measurement 
be the superposition of two packets ${\psi}_{1}$ and ${\psi}_{2}$, each 
hitting one detector. The evolution of the wave function of the system 
${\cal S} + {\cal B}$  during the measurement is ruled be the Schr\"odinger 
equation and is of the type (3), \iec, in the present notation, 
\begin{equation} 
\mbox{$\;\;\;$} 
{\Psi}^{\rm i}=\left( c_{1}{\psi}_{1}(x) + c_{2}{\psi}_{2}(x) \right)
{\beta}^{\rm i}(\mbox{\boldmath $y$})\stackrel{\rm S.E.}{\llongrightarrow}
{\Psi}^{\rm f}= c_{1}{\psi}_{1}(x) {\beta}_{1}^{\rm f}(\mbox{\boldmath $y$})
 + c_{2}{\psi}_{2}(x) {\beta}_{2}^{\rm f}(\mbox{\boldmath $y$})
\end{equation}
where we have omitted to indicate the trivial time evolution bringing the 
particle through the detectors. In (21) ${\beta}^{\rm i}$, 
${\beta}_{1}^{\rm f}$ and ${\beta}_{2}^{\rm f}$ are the wave functions of 
$\cal B$ before the measurement, after the 
\begin{figure}[h]
\bigskip
\centerline{\figura}
\smallskip
\begin{quote}
{\footnotesize\dicitura}
\end{quote}
\end{figure}
measurement if the particle wave function were ${\psi}_{1}$ alone, and after 
the measurement if the particle wave function were ${\psi}_{2}$ alone, 
respectively. ${\beta}_{1}^{\rm f}$ and ${\beta}_{2}^{\rm f}$ are strongly
peaked around $z = z_{1}$ and $z = z_{2}$, respectively. The evolution of the 
configuration of ${\cal S} + {\cal B}$ is ruled by the guidance equation 
and can be written 
\begin{equation}
\overline{x}, \mbox{\boldmath $y$}^{\rm i} 
~~~\stackrel{\rm G.E.}{\llongrightarrow}~~~
\left\{ \begin{array}{ll}
\overline{x}', \, {\overline{z}}^{\rm f} = z_{1},\, 
\ldots & \mbox{if $\overline{x}$ lies within ${\psi}_{1}$},\\
\smalldisplayskip
\overline{x}', \, {\overline{z}}^{\rm f} = z_{2},\, 
\ldots & \mbox{if $\overline{x}$ lies within ${\psi}_{2}$,}
\end{array} \right.
\end{equation}
where $\overline{x}$ and $\overline{x}'$ are the initial and final positions 
of the particle $\cal S$ for the particularly considered member of the 
ensemble (the prime is intended to describe the effect of the diffusion 
present in \eq6), $\mbox{\boldmath $y$}^{\rm i}$ is the initial configuration 
of $\cal B$, and ${\overline{z}}^{\rm f}$ is the final value of the pointer 
variable of the detectors. The wave function ${\Psi}^{\rm f}$ is a 
superposition of two terms disjoint in configuration space. After the 
measurement, each member of the ensemble of systems ${\cal S} + {\cal B}$ is 
described by ${\Psi}^{\rm f}$ (the same for all members) and by the 
coordinates $\overline{x}', {\overline{z}}^{\rm f}, \ldots$ (particular values 
for a particular member). The wave function ${\Psi}^{\rm f}$ is no more 
inconceivable because it is only a part of the complete state and CSR is 
satisfied. It is obvious that, if the measurement is repeated by a second 
apparatus similar to $\cal A$ possibly contained in $\cal B$, corresponding 
results are obtained for each member of the ensemble. 

As already noted, the two terms in the wave function ${\Psi}^{\rm f}$ 
immediately after the measurement have supports disjoint in configuration 
space. Because of the enormous complexity of the system $\cal B$ (see \eg 
\cite{Bohm-Hiley}), this condition will remain true for all subsequent times 
with overwhelming probability, in spite of any possible attempt to rejoin the 
two terms. This circumstance, in conjunction with the assumption that any 
measurement is ultimately a position measurement, is the form taken by 
effective incoherence in hidden--configuration theories and it ensures that 
the statistical predictions of these theories as regards any possible 
subsequent measurement coincide practically with those of standard quantum 
mechanics. 

One further point remains to be discussed. The condition $\rho = |{\Psi}|^{2}$ 
is satisfied for the originally considered ensemble $\mbox{\boldmath $E$}$ 
after the measurement as it is before the measurement, according to the 
discussion in the previous section. However, after the measurement, 
$\mbox{\boldmath $E$}$ can be split into two parts, $\mbox{\boldmath $E$}_1$ 
and $\mbox{\boldmath $E$}_2$, depending on whether the particle has been found 
in one detector or the other. Considering \eg $\mbox{\boldmath $E$}_1$, the 
systems contained in it are described, according to the discussion above, by 
the wave function $\Psi=c_1\Psi_1+c_2\Psi_2 = c_{1}{\psi}_{1}(x) 
{\beta}_{1}^{\rm f}(\mbox{\boldmath $y$}) + c_{2}{\psi}_{2}(x) 
{\beta}_{2}^{\rm f}(\mbox{\boldmath $y$})$, but the probability distribution 
is $\rho = |{\Psi_1}|^{2}$ corresponding to $\Psi_1$ alone. Therefore, the 
ensemble $\mbox{\boldmath $E$}_1$, whose consideration is perfectly 
legitimate, violates the condition $\rho = |{\Psi}|^{2}$. This fact does not 
give rise to contradictions, because the condition $\rho = |{\Psi_1}|^{2}$ for 
the systems in $\mbox{\boldmath $E$}_1$ remains true when time passes even 
maintaining that the wave function is the whole $\Psi$. Indeed, the dynamics 
of each system ${\cal S} + {\cal B}$ is specified by the initial value of the 
wave function $\Psi$ together with the Schr\"odinger equation governing its 
evolution and by the initial value of the configuration $(x,\mbox{\boldmath 
$y$})$ together with the guidance equation governing (stochastically, if 
$\alpha \ne 0$) its evolution, the drift appearing in the latter equation 
being in turn determined by $\Psi$. Assumptions on the distribution $\rho$ of 
configurations at a given time in a given ensemble have nothing to do with the 
fundamental dynamics and only serve the practical purpose of making 
statistical predictions about the ensemble. The role of such assumptions is 
similar to the role of the choice of statistical ensembles in classical 
statistical mechanics. If an ensemble is put together using informations about 
configurations, as in the case of the ensemble $\mbox{\boldmath $E$}_1$ above, 
it is natural and obligatory taking into account such informations assuming 
$\rho = |{\Psi_1}|^{2}$. Then the agreement with the known phenomenolgy 
described by standard quantum mechanics is ensured by the fact that this 
condition is conserved in time and the consistency of the theory by the fact 
that such a conservation holds even using as the wave function the whole 
$\Psi$, because $\Psi_1$ and $\Psi_2$ remain forever separated in 
configuration space and because of the remark at the end of the previous 
section. In conclusion, each individual physical system is described by the 
values of the coordinates of all particles in the system and by a wave 
function which never suffered any reduction. 

\section{Conclusion}

The content of the previous sections is to be compared with that of similar 
works. As already stated in the Introduction, the theory of quantum 
measurement in the framework of Bohmian mechanics has already been developed 
along the same lines followed here by Bell \cite{Bell,Bell1}, Bohm and Hiley 
\cite{Bohm-Hiley}, and D\"{u}rr, Goldstein and Zangh\`{\i} \cite{DGZ}. It has 
also been retraced on the basis of the same simple example used in Section 4 
by one of the present authors \cite{Rimini}. 

The theory of quantum measurement in the framework of Nelson's stochastic 
mechanics has been worked out in a paper by Blanchard, Cini and Serva 
\cite{BCS}, and, previously, by Goldstein \cite{Goldstein}. The arguments used 
by Blanchard, Cini and Serva are very similar to those developed here, but 
they are presented in a form which implies that the use of Nelson's stochastic 
mechanics is crucial to get a satisfactory description of measurement. We 
think, on the contrary, that what is crucial is solely the addition to the 
wave function of a new element of description (the configuration), which 
allows to satisfy the proposition referred to above as Common Sense 
Requirement. As we have seen, this is possible for all formulations of quantum 
mechanics of the family introduced in Section 3. Even though the paper of 
Goldstein refers only to Nelsonian mechanics, that point is clear in his 
presentation.\footnote{There is no relation between the theory of measurement 
in the framework of hidden configuration theories presented in Section 4 and 
the reduction-like processes hypothesized by Guerra \cite{Guerra} in the 
framework of discrete generalization of stochastic variational principles. 
Even though it is not explicitly stated, the theory described there concerns 
the quantum system alone (the system we have called ${\cal S}$), while in 
Section 4 here the system is ${\cal S} + {\cal A} + {\cal E}$, and in this 
sense a theory of measurement is developed. As we have seen, in such a context 
no reduction has to be introduced, and a satisfactory description of the 
measurement process is nevertheless obtained. May be that the reduction--like 
process, which is possibly present in the formalism of Guerra, can be used to 
describe the effective reduced dynamics for the quantum system $\cal S$ 
alone.} 

The unified treatment of the measurement process in Section 4 for all 
formulations of the family described in Section 3 helps in the clear 
identification of the true crucial point allowing the elaboration of a 
satisfactory theory of quantum measurement. Furthermore it seems to us that, 
at this point, one should address the question whether the introduction of a 
diffusion is really worthwhile, having realized that it is not essential to 
solve the problems of precision and consistency of quantum 
mechanics.\footnote{Nevertheless, it may reveal useful to exploit the well 
developed mathematical methods of probability theory and stochastic processes 
in quantum mechanics, in particular in its path integral formulation.}

\end{document}